\documentclass[12pt]{iopart}
\usepackage[utf8]{inputenc}
\usepackage{latexsym}
\usepackage{color}
\usepackage{graphicx}
\usepackage{epsfig}
\usepackage{amsopn}
\usepackage{amsfonts}
\usepackage{amssymb}
\usepackage{subfigure}
\usepackage{braket}
\usepackage{epstopdf}
\usepackage{braket}
\usepackage{nicefrac}
\usepackage{textgreek}

\begin{document}

\title{Fabrication of $^{15}\textrm{NV}^{-}$ centers in diamond using a deterministic single ion implanter}

\author{Karin Groot-Berning} \address{QUANTUM, Johannes Gutenberg-Universit\"at Mainz, Staudinger Weg 7, 55128 Mainz, Germany}
\ead{karin.groot-berning@uni-mainz.de}
\author{Georg Jacob} \address{Alpine Quantum Technologies GmbH, Technikerstrasse 17 / 1, 6020 Innsbruck, Austria}
\author{Christian Osterkamp}\address{Institut f{\"u}r Quantenoptik, Universit\"at Ulm, Albert Einstein Allee 11, 89081 Ulm, Germany}
\author{Fedor Jelezko}\address{Institut f{\"u}r Quantenoptik, Universit\"at Ulm, Albert Einstein Allee 11, 89081 Ulm, Germany}
\author{Ferdinand Schmidt-Kaler} \address{QUANTUM, Johannes Gutenberg-Universit\"at Mainz, Staudinger Weg 7, 55128 Mainz, Germany}
\address{Helmholtz-Institut Mainz, 55128 Mainz, Germany}

\begin{abstract}
Nitrogen Vacancy (NV) centers in diamond are a platform for several important quantum technologies, including sensing, communication and elementary quantum processors. In this letter we demonstrate the creation of NV centers by implantation using a deterministic single ion source. For this we sympathetically laser-cool single $^{15}\textrm{N}_2^+$ molecular ions in a Paul trap and extract them at an energy of 5.9\,keV. Subsequently the ions are focused with a lateral resolution of 121(35)\,nm and are implanted into a diamond substrate without any spatial filtering by apertures or masks. After high-temperature annealing, we detect the NV centers in a confocal microscope and determine a conversion efficiency of about 0.6\,$\%$. The $^{15}\textrm{NV}$ centers are characterized by optically detected magnetic resonance (ODMR) on the hyperfine transition and coherence time. 
\end{abstract}

\maketitle

\section{Introduction}
\subsection{Motivation}
In the past two decades research on single nitrogen vacancy (NV) centres in diamond has undergone a dramatic progress. Since the first observation of single NVs with a confocal microscope and the demonstration of optically detected magnetic resonance (ODMR) \cite{gruber1997scanning}, there have been numerous experiments which show a multitude of applications covering many fields such as metrology and quantum information processing. Among these applications are quantum sensors for magnetic and electric fields on the nanometer length-scale as well as microwave sensors \cite{Muller2014}. 

A well established method to create NV centers is the implantation of nitrogen ions with subsequent annealing of the sample \cite{meijer2005generation}. This approach is especially beneficial in cases where a precise placement of NVs is necessary. Typically, these are applications where the NVs are placed within dedicated structures in order to couple them to light fields \textit{e.g.} inside a photonic waveguide structure \cite{riedrich2015nanoimplantation,schroder2017scalable} or a solid immersion lens \cite{hadden2010stronglyenhanced}. In these applications, implantation allows for circumventing the necessity of manufacturing such structures around pre-existing NV centers. The requirements on the resolution thereby, is given by the wavelength of the optical fields and thus is in the order of less than 100\,nm. To this date, various techniques have been proposed and developed to reach that aim. For example, nanofabricated masks or pierced AFM tip which provide apertures \cite{Pezzagna2011, Meijer2008}. Another approach is using a focused ion beam with the respective resolution. This circumvents the need for employing a mask or an AFM tip near the focal plane \cite{schroder2017scalable}. However, these techniques are using stochastic sources, limiting the applications to cases where single NV devices can be post selected depending on whether an ion was implanted or not. This rules out applications which rely on coupling NV centers via their mutual dipolar magnetic interaction which is on the scale of a few tens of nanometers. Such a coupling between two NVs has been demonstrated in various experiments \cite{neumann2008multipartite,dolde2013room, Yamamoto2013}. Scalable use of this resource e.g. for creation of entanglement in the context quantum information processing calls for deterministic placement of single NVs with nanometer resolution. The need for arrays of single NVs at nanometer accuracy is even more important in view of building a scaled-up quantum processor or simulator based on this solid state platform \cite{wu2019,abo2019,bradley2019}.

To this goal we implement an intrinsically deterministic ion source by repetitively loading a single laser-cooled nitrogen molecular ion into a linear Paul trap and launch it from there. The laser cooling provides a small phase space occupation of the generated beam in both, the transversal and longitudinal direction. The former allows for tight focusing without need for spacial filtering which would destroy the deterministic property of the source. The latter results in a low energy dispersion, important for avoiding chromatic aberration when focusing by electric field lenses. Additionally, our method uses singly charged ions at energies lower than 10\,keV, unlike methods that rely on the detection of single ion impact events \cite{jamieson2020}. The low energy implantation reduces position uncertainty due to straggling and surface destruction of the bulk diamond. Likewise the very same apparatus allows for transmission imaging of the substrate using single extracted calcium ions \cite{jacob2016transmission}. This provides a precise referencing and positioning of the dopants with respect to transmissive markers, free of parallax errors. Minimal charging and irradiation of the diamond substrate is ensured by using single ions for imaging. In this paper we present a proof of principle experiment which demonstrates the creation of NV centers with high resolution by focusing a nitrogen beam generated by this source. Although the deterministic production of single NV centers with this method is currently severely limited by the creation yield, additional measures - such as co-implantation of Sulfur \cite{lue2019}, surface termination \cite{hauf2011chemical}, diamond overgrowth \cite{lesik2016production} and electron irradiation \cite{schwartz2012effects} - could realize such a truly deterministic creation process in the future. Also, the apparatus can be used as a deterministic source of any atomic and molecular ions and we envision for the future co-implanting of other ion species, e.g. $^{13}$C$^+$ ions, to tailor the spin-environment of the NV center.

\section{Experimental Apparatus and Procedures}

\begin{figure}[htb]
\centering
\includegraphics[scale=0.85]{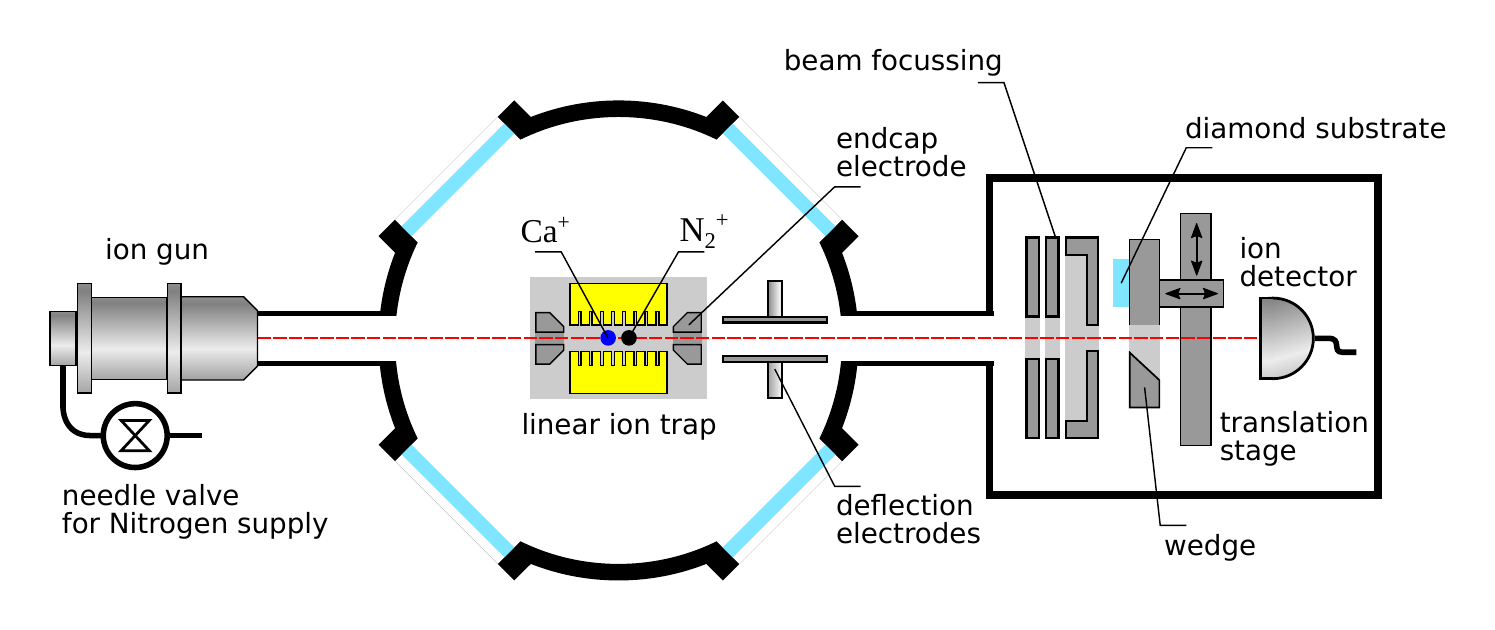}
\caption{Sketch of the experimental setup: The ion gun serves as a source for molecular $^{15}$N$_2^+$ ions, which are loaded into the linear Paul trap together with atomic Ca$^+$ ions from an oven.  A Wien filter after the ion gun (not shown) may be  used. The ion extraction from the Paul trap is acieved by applying a voltage pulse to a pierced endcap electrode. A pair of deflection electrodes are used for beam steering, and the beam is focused with an Einzel-lens. A three-dimensional nano-translation stage with the probe holder and an optional wedged blade can be used for ion beam characterization together with the secondary electron multiplier (SEM).
}
\label{pic:setup}
\end{figure}

A linear Paul trap acts as an ultracold ion source \cite{karin2019,Meijer2006concept,schnitzler2009deterministic,izawa2010controlledExtraction}. The trap consists of four gold coated alumina chips mounted in a X-shaped arrangement, see Ref.~\cite{karin2019} and Fig.~\ref{pic:setup} for details. One pair of diagonally opposing chips are supplied with RF voltage giving rise to a radial confinement of the ions. The chips of the other pair is segmented into 11 electrodes. These segments allow for shaping the axial potential by applying DC voltages. Along the axial direction, the trap is encapsulated by two pierced end-caps with a length of 10\,mm, allowing for the extraction of the ions by switching them to high voltage.

Calcium atoms are provided by an oven which is directed towards the center of the trap. Inside the trapping volume the atoms are photo-ionized, trapped and laser cooled with light at 397\,nm on the S$_{1/2}$ to P$_{1/2}$ dipole transition. Trapped ions are detected and  automatically counted, by imaging their fluorescence onto an EMCCD-camera\,\footnote{Andor iXon X3, \textit{DU-860E-CS0-UVB}, Andor Technology, Belfast, Northern Ireland}. The loading of a predefined number of calcium ions is accomplished by an automated procedure: First, a random number of ions is trapped, cooled and counted from the camera image. If necessary, ions are removed by lowering the axial trapping potential with a predefined voltage sequence. Subsequently, the success of this sequence is evaluated by counting the number of ions again and in case of discrepancy, the procedure is repeated.

Ion species other than calcium are loaded by means of a commercial ion gun\,\footnote{Ion Source \textit{IQE 12/38}, SPECS Surface Nano Analysis GmbH, 13355 Berlin, Germany} from  gaseous sources as well as solid sources. The remainder of this paper is solely concerned with making use of the gaseous source. If the reader is interested in single-ions-on-demand from solid sources we refer to examples for praseodymium and thorium~\cite{karin2019,karin2019b}. We control the flux from the isotopically pure $^{15}$N$_2$ gas source \footnote{Gas source $^{15}$N$_2$, 364584-1L-EU, 98$\%$ 15N, Sigma-Aldrich} into the ion gun volume with a needle valve. Here, the molecules are ionized by electron impact. Subsequently the ions are extracted with typically 500\,eV and are collimated to a beam which is guided and focused onto the hole of one of the trap endcaps. Deflection electrodes allow for blanking of this beam \textit{i.e.} switching the loading of the trap on and off. For the gaseous source it is sufficient to direct the ion beam into the trap for a fixed period of time. Within an average loading time of about 30 seconds we are trapping one $^{15}$N$_2$ ion. We conjecture a loading mechanism facilitated by the modulation of the axial trapping potential due to the RF drive. This allows ions to enter the trap at the lower turning point of the axial potential modulation and keep them confined sufficiently long for sympathetic cooling such that their energy is reduced below the trap depth.

The nitrogen molecular ions are sympathetically cooled via their Coulomb interaction with the laser cooled calcium ions. A voltage sequence, similar to the aforementioned sequence for the loading of a given number of calcium ions, is applied in order to prepare a crystal consisting of exactly one calcium and one nitrogen molecular ion. The existence of a trapped nitrogen ion is detected by a shift of the calcium ion from its former equilibrium position on the camera image to the side, whereas the other ion does not emit light, thus coined briefly $\textit{"dark ion"}$.

The deterministic source is implemented by extracting the single ion(s) from the trap.
The accelerating electric field is provided by applying high voltages of up to $-3~$kV to one of the pierced endcaps. The ion kinetic energy is doubled by switching the voltage to a positive value while the ion is inside the endcap hole. Alignment and scanning of the ion beam is accomplished by two pairs of deflection electrodes which are placed along the ion pathway. Depending on the charge to mass ratio, the dark ion is either faster or slower than the calcium ion. This fact can be harnessed to separate the calcium. In case of a higher charge to mass ratio compared to calcium, the dark ion \textit{e.g.} $^{15}$N$_2$ will arrive earlier at the endcap. At the moment the dark ion is inside the endcap, the voltage of the endcap is switched to a positive value, deflecting the calcium ion back when approaching the endcap. In the case where the charge to mass ratio is lower, the dark ion will leave the endcap later than the calcium ion. Switching the voltage of the endcap to a positive value is performed when the calcium ion has already left the endcap, resulting in a lower energy compared to the dark ion. Because of this, the deflection electrodes act differently on the two ion species, resulting in a separation of the calcium ion.

\begin{figure}[htb]
\centering
\includegraphics[scale=0.4]{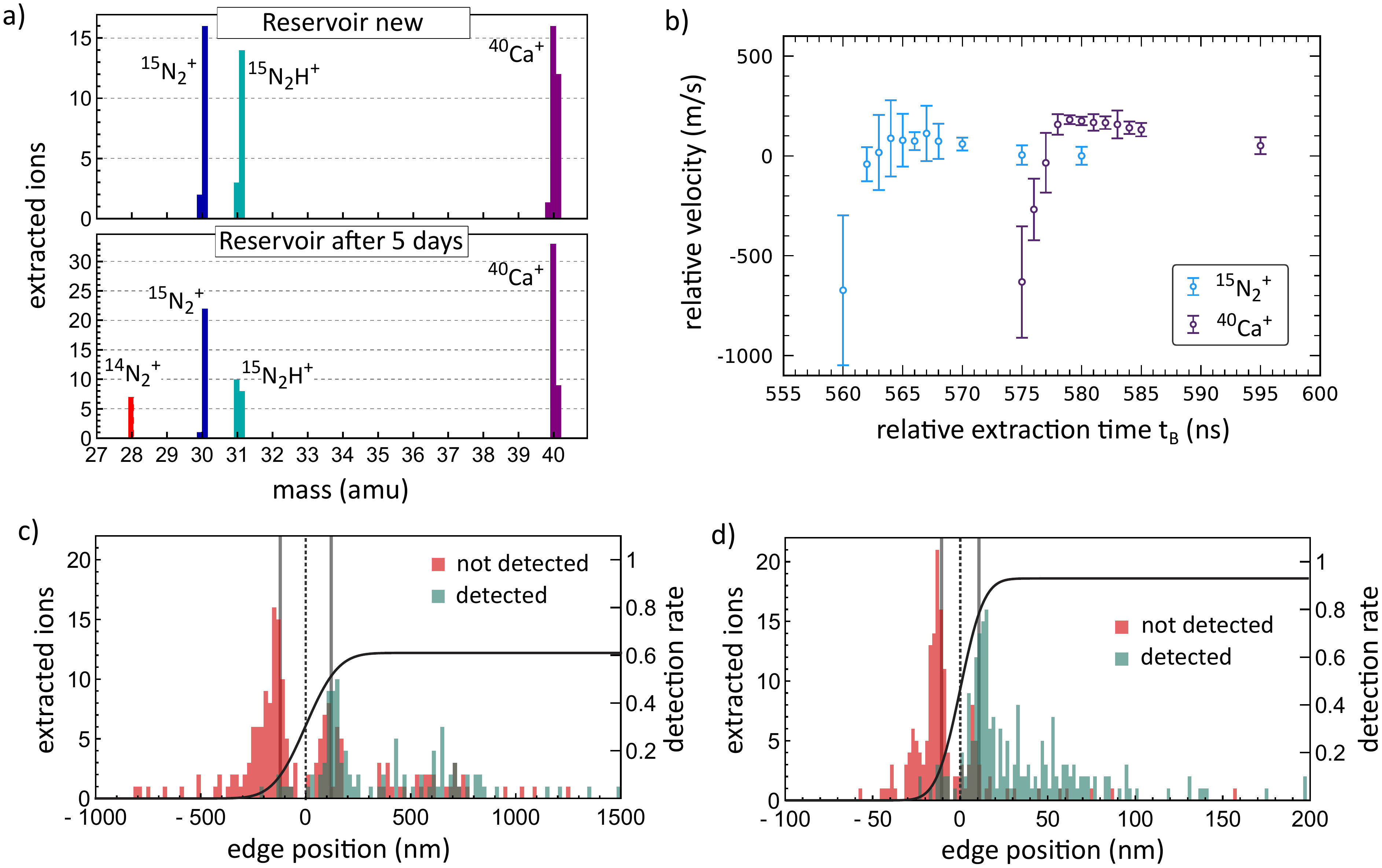}
\caption{Ion beam characterization: We employ the SEM detector and remove the diamond sample from the ion pathway.
a) TOF signal in the SEM from single $^{15}$N$_2^+$, $^{15}$N$^{14}$N$^+$,  and $^{14}$N$_2^+$ ions after a 0.428~m travel. After carefully purging all tube connections between $^{15}$N$_2$ gas bottle and ion gun we found no indication for $^{14}$N$_2^+$ ions. However, after a waiting time of 5 days, the histogram shows a peak at mass 28, which was assigned to the isotope $^{14}$N.
b) Velocity variations of $^{40}$Ca (purple) and $^{15}$N$_2$ (blue) ions as a function of delay between RF and DC switching. Ions with a mass of 40 and 30, respectively, arrive at different times at the position of the endcap hole, and explore different phases of the RF drive field.
c) Focus spot determination of Ca ions, measured by sweeping the wedge into the beam and observing the partial blocking, which is detected from the SEM counts. For $^{40}$Ca$^+$ ions we obtain a $\sigma$ = 11(2)\,nm in x-direction. The data acquisition time for the extraction of 380 ions was approximately 17~min, which corresponds to a loading rate of 22.4\,ions/min.
We conjecture thermal and electrical drifts as well as mechanical vibrations as major source of beam pointing fluctuations.
d) Focus spot determination of nitrogen molecular ions yielding $\sigma$ = 121(35)\,nm in x-direction. The measuring time for the extraction of 289 nitrogen ions takes about 117\,min, which results in a loading rate of 2.5\,ions/min.
}
\label{pic:beam}
\end{figure}

For an unambiguous determination of the dark ion species, i.e. to identify the trapped particle as a successfully loaded $^{15}$N$_2^+$ molecular ion, we extract theions and detect them after a flight of 428\,mm in length using a secondary electron multiplier with an efficiency of 96$\pm$2\,$\%$. During ion extraction, we switch off the RF amplitude, to avoid shot-to-shot modifications of the ion trajectory. These modifications originate from the time dependent electric fields of the RF drive at the vicinity of the endcap in combination with a timing jitter of the exact onset of the  extraction voltage. We determine the mass of the extracted ions, from time-of-flight (TOF) measurements, discriminating between $^{15}$N$_2^+$, $^{15}$N$^{14}$N$^+$, and $^{14}$N$_2^+$, see Fig.~\ref{pic:beam}(a). The trigger for the HV switching and RF switching is chosen such that the velocity modification is minimal, see Fig.~\ref{pic:beam}(b). Prior to extraction, we arrange the order of the ions in the linear crystal, such that the lighter nitrogen ion is ahead. This prevents a Coulomb interaction of the nitrogen ion with the havier and thus slower calcium ion. The ordering of ions is achieved by briefly melting and then recrystallizing the crystal, followed by a check of the right ion order from an EMCCD image of the fluorescence. In future, we may apply deterministic and fast swapping, as demonstrated experimentally~\cite{kau2017}. We focus the ion beam by an electrostatic lens, see \cite{jacob2016transmission} for details. To optimize and finally characterize the spot size, we sweep a mechanical wedge into the focus and find $\sigma\textsubscript{Ca}$ = 11(2)~nm, and $\sigma\textsubscript{N2}$ = 121(35)~nm, respectively. We conjecture that long term drifts are affecting the spot size of the N$_2^+$, as the data acquisition time for the Ca$^+$ spot measurement is about 2 orders of magnitude faster. Sympathetic cooling of a dark ion of mass u=30 with $^{40}$Ca$^+$ is expected to work efficiently, since the masses are about equal in the mixed crystal and thus the coupling of radial vibrational degrees of motion is still sufficient~\cite{wuebbena2012}. 

\section{Sample Preparation and Implantation}
\begin{figure}[htb]
\centering
\includegraphics[scale=0.4]{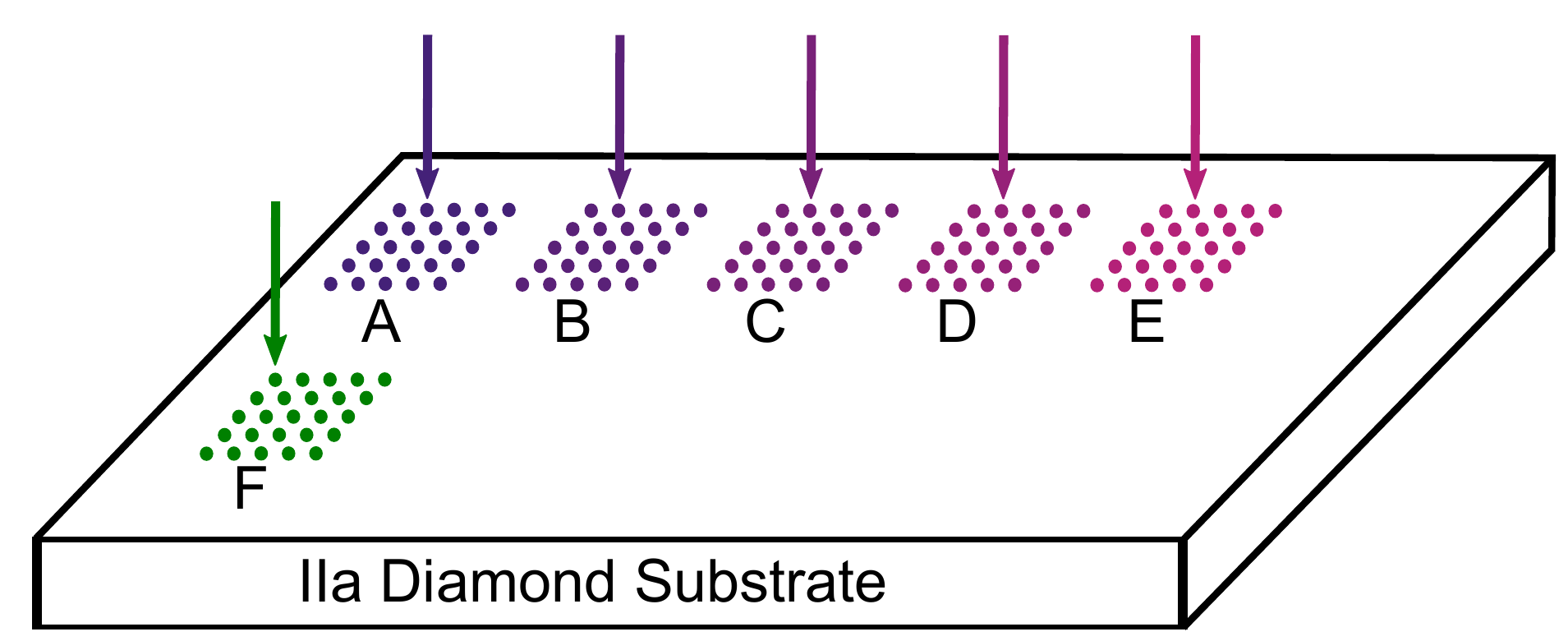}
\caption{Implantation pattern with different ion quantities $\kappa$ per spot which are separated by 2$\,\mu$m. Regions A-E (purple) were implanted with $^{15}$N$_2$$^+$ ions whereas region F (green) was implanted with $^{15}$N$_2$$^+$ and Ca$^+$ ions. A: $\kappa_A=$20$\,\frac{\mathrm{ions}}{\mathrm{spot}}$, B: $\kappa_B=$10$\,\frac{\mathrm{ions}}{\mathrm{spot}}$, C: $\kappa_C=$4$\,\frac{\mathrm{ions}}{\mathrm{spot}}$, D: $\kappa_D=$2$\,\frac{\mathrm{ions}}{\mathrm{spot}}$, E: $\kappa_E=$1$\,\frac{ion}{spot}$, F: $\kappa_F=$20$\,\frac{\mathrm{ions}}{\mathrm{spot}}$}
\label{pic:pattern}
\end{figure}

Before any implantation was performed, the diamond host sample, a commercial type IIa electronic grade diamond from supplier Element Six, was investigated with a home-built confocal microscope setup which is described in Sect. 4. Stable NV$^-$~centers were found in a concentration of 1 NV$^-$~per 100 \textmu m$^2$ (or 0.27 parts per trillion, ppt). After careful cleaning, we fix the sample on the three-axis nano-positioner in the UHV setup, see Fig.~\ref{pic:setup}. The diamond was exposed to accelerated focused nitrogen ions of isotope $^{15}\textrm{N}_2^+$. Since each single nitrogen ion is trapped and identified prior to implantation, the dose can be controlled at a single ion level. A 5x5 pattern was implanted with a distance of 2$\,\mu$m between the single implantation spots by moving the sample with the nano-positioner. The dose $\kappa$ can be adjusted freely and was varied from one to 20 ions per spot, see  Fig.~\ref{pic:pattern} for details. According to SRIM simulations the used 3\,keV implantation energy per atomic ion corresponds to a penetration depth of 4.2\,nm and therefore the resulting NV centers can be considered as shallow \cite{Ziegler2012}.

The implanted sample was acid cleaned in a mixture of sulfuric, nitric and perchloric acid (ratio 1:1:1) which was heated to 130$\,^{\circ}$C for 2 hours, in order to remove any dirt from the surface. Especially any graphitic layer which could be produced during the ion bombardment is removed by this treatment.

A successive annealing procedure under ultra-high vaccum was  subsequently performed in order to activate NV centers. During this process, vacancies (empty lattice sites) that are created due to collisions during the implantation process, become mobile and pair with implanted nitrogen atoms forming stable nitrogen-vacancy centers \cite{Naydenov2010}. The temperature was set to 250$\,^{\circ}$C for one hour to keep a high vacuum during the annealing process. Then, the temperature was ramped up to 900$\,^{\circ}$C and hold for another two hours, with a vacuum of 10$^{-7}\,$mbar. Finally, the system was cooled down. Another acid boiling step was included to ensure the oxygen termination of the diamond surface, which is essential for the preservation of the NV$^-$ charge state of shallow NVs \cite{Hauf2011, Osterkamp2013}.

\section{NV Characterization}

\begin{figure}[htb]
\centering
\includegraphics[scale=0.7]{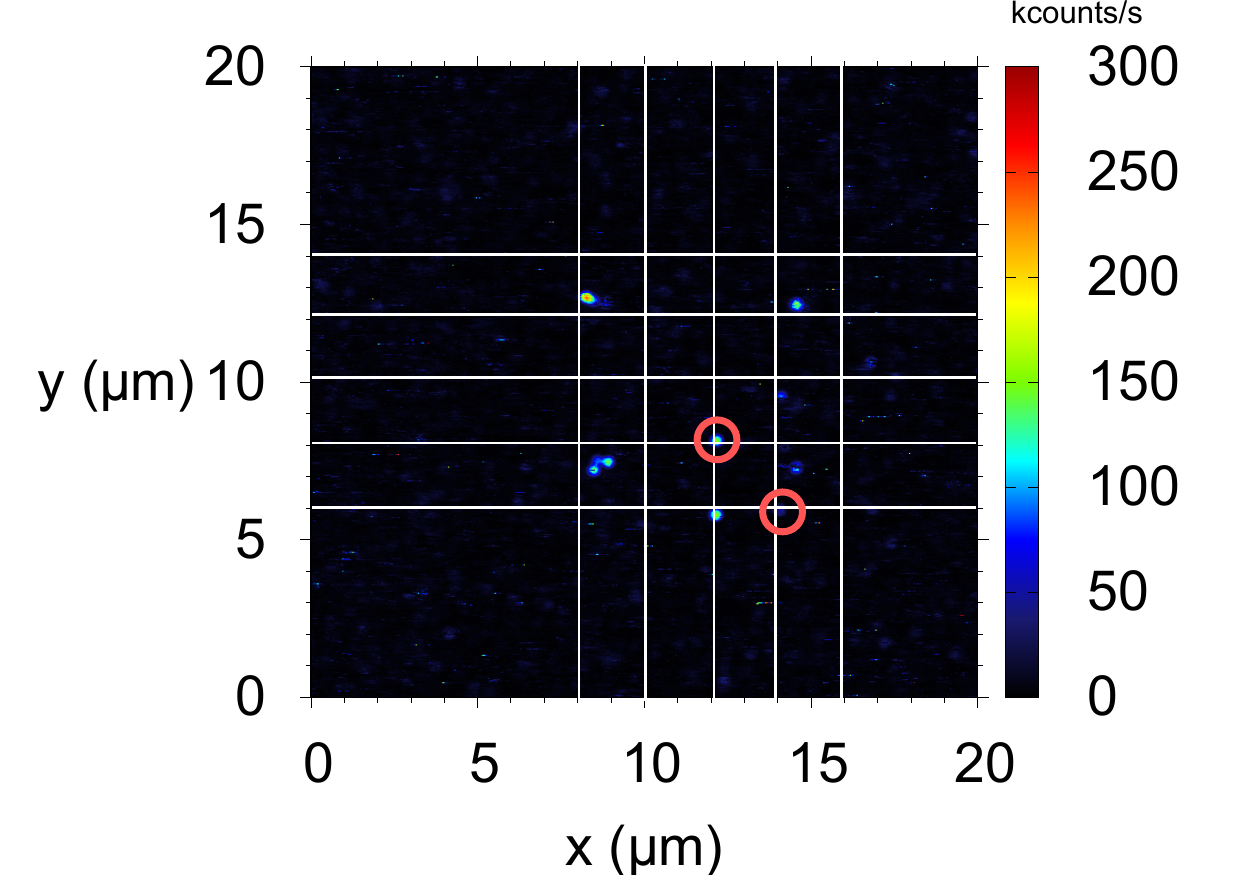}
\caption{Confocal image of the implanted region A. A grid pattern of 2$\,$\textmu m distance is displayed to guide the eye. The bright spots correspond to NV centers present in the diamond. The presence of six NV centers on implanted sites was confirmed of which three were found in neutral charge state and two with the implanted nitrogen isotope $^{15}$N.}
\label{pic:confocal}
\end{figure}

\begin{figure}[htb]
\centering
\includegraphics[scale=0.5]{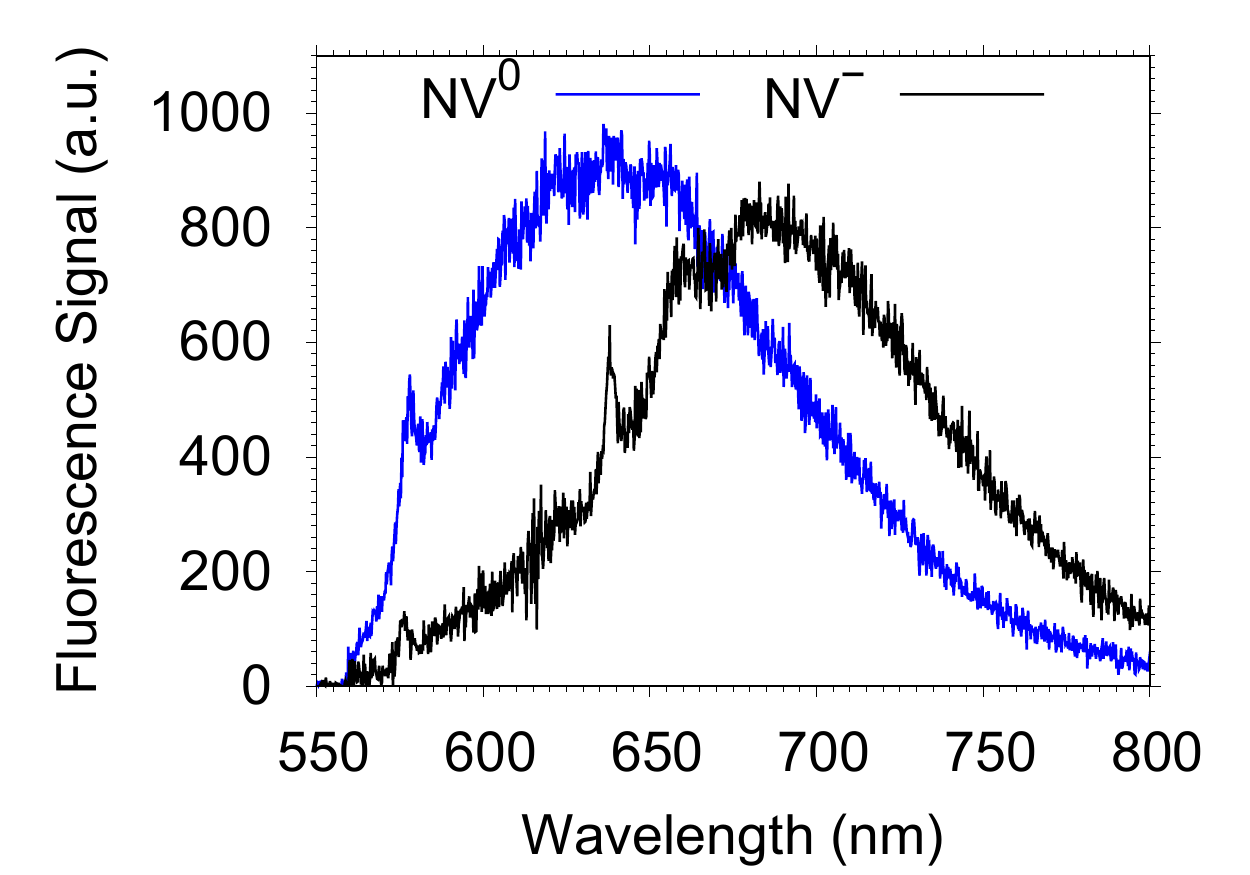}
\caption{NV$^-$~ (black) and NV$^0$~ (blue) fluorescence spectra. The two different charge states are optically distinguishable since NV$^-$ shows a ZPL of 637$\,$nm whereas the one for NV$^0$ lies at 575$\,$nm.}
\label{pic:spectrum}
\end{figure}

\begin{figure}[htb]
\centering
\includegraphics[scale=0.5]{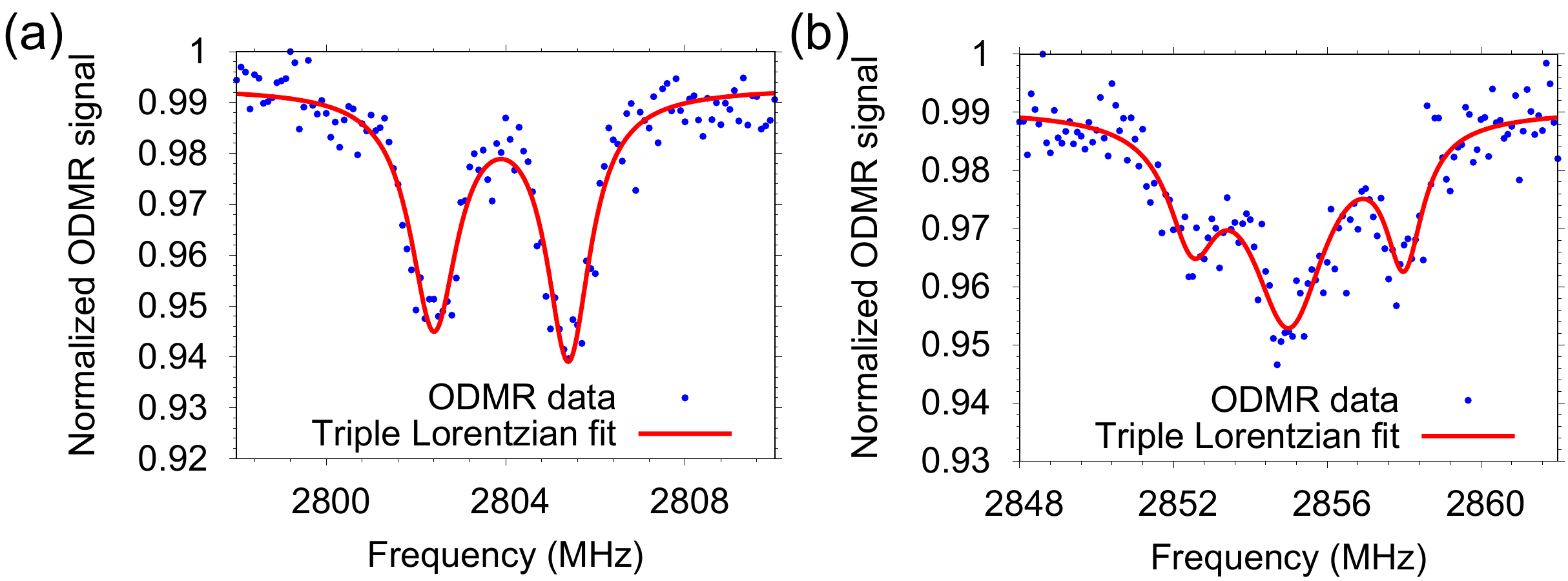}
\caption{Pulsed ODMR spectra of the NV$^-$~centers created by the deterministic implantation method. (a) Two lines, separated by 3.1$\,$MHz correspond to the hyperfine coupling to $^{15}$N nitrogen nucleus. This spectrum is proof that the implantation method is working as desired. (b) Hyperfine coupling of the NV's electronic spin to the nitrogen nuclear spin reveals the presence of a $^{14}$NV$^-$, since the characteristic three line structure becomes visible, where the lines are split by 2.2$\,$MHz.}
\label{pic:podmr}
\end{figure}

Confocal imaging was performed on a home-built confocal microscope consisting of a 518$\,$nm diode laser, a movable micrometer stage, an oil immersion objective (NA=1.4) and an avalanche photo diode. A lateral x-y-scan of the diamond sample can be seen in Fig.~\ref{pic:confocal}.
Clearly visible are spots with high count rate (bright spots) coming from the NV center fluorescence to which the setup is optimized by an optical filtering system (excitation: 535/20BP, detection: 560LP). In total, six stable spots can be observed in region A, showing the typical fluorescence spectrum of NV centers, see Fig.~\ref{pic:spectrum}.
Some of them are present in the NV$^0$ charge state with no accessible electron spin and therefore cannot be used for electron spin resonance (ESR) measurements. The fluorescence spectrum of NV$^0$ shows a zero-phonon-line (ZPL) at 575$\,$nm with a band to higher wavelengths. The ones which show a ZPL at 637$\,$nm (as marked with red circles in figure \ref{pic:confocal}) are in the NV$^-$ configuration and can be further investigated. These NV centers show fluorescence intensities which are typical for the presence of a single emitter. One of the key features of NV centers is that the fluorescence level itself depends on the state of its electron spin. Therefore the ESR transition of the NV center can be determined in an optically detected magnetic resonance (ODMR) experiment. To this end, the NV is continuously irradiated with a green laser while a microwave with varying frequency is applied. In the absence of a magnetic field a zero field splitting of 2.87$\,$GHz confirms the presence of an NV$^-$ center.
Furthermore, it is also possible to determine whether a center has formed due to the presence of nitrogen isotopes 14 or 15. To do this, the mutual effects described above must be minimized in order to obtain a resonance dip linewidth with which the hyperfine coupling to the nitrogen nucleus can be measured. A technique which offers this requirements is the pulsed ODMR technique where the microwave sweep is performed in a pulsed way inverting the electron spin states. Clearly visible is the separation of 3.1$\,$MHz of the ODMR measurement in Fig.~\ref{pic:podmr}, which is related to presence of a $^{15}$N nuclear spin.

A nitrogen 14 nuclear spin would be imprinted as a triplet in the ESR signal (I=1), where the lines are separated by 2.2$\,$MHz \cite{Rabeau2006}. Only two of the found NVs show the hyperfine splitting of $^{15}$N. Another one did show the triplet feature and the rest were present in the wrong charge state, so no ODMR signal was observed. The presence of $^{14}$NV can be explained by the following. During the implantation of $^{15}$N$^+$ ions, collision to carbon atoms cannot be avoided even at low implantation energies, therefore vacancies are created when the ions penetrate into the diamond. The number of additional vacancies exceeds the number of $^{15}$N ions. Also, if the Ca$^+$ is not blanked out, vacancies might be generated by its impact. The generated vacancies may recombine with natural $^{14}$N atoms in the diamond substrate present in ppb concentartion in electronic grade CVD diamond crystals~\cite{Yamamoto2013}. Another explanation is that not solely $^{15}$N ions were implanted but accidentally also $^{14}$N ions, as the nitrogen reservoir was not immediately purged and refilled before implantation, see Figure \ref{pic:beam}a, and the Wien filter selectivity might not be sufficient. However, the origin of the $^{14}$NV center cannot be clarified entirely. 


The NV creation yield of our implantation is estimated by dividing the number of observed centers by the number of implanted nitrogen ions. We calculate a NV creation yield of 0.6\% for region F. This is consistent with previous reports \cite{Pezzagna2010} at this low energies of implanted N ions.

For the implantation attempts without the co-implantation of Ca$^+$, NVs are successfully created only in region A (highest dose). This suggests that the number of created vacancies, at very low doses, is limiting the creation yield. Note, that here all NVs found there stem from $^{15}$N$^+$ implantation events. 
\begin{figure}[htb]
\centering
\includegraphics[scale=0.5]{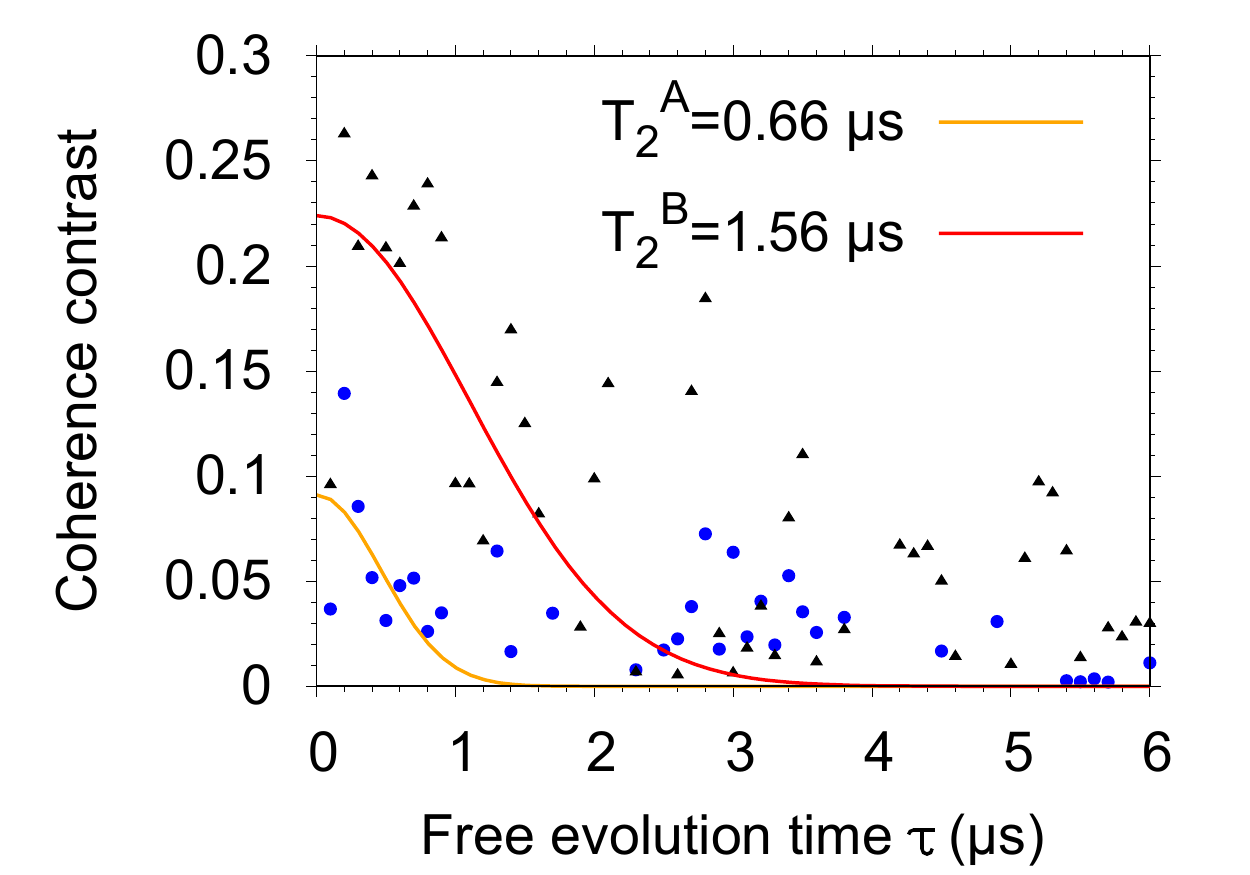}
\caption{Hahn echo measurements performed on the successfully implanted $^{15}$NV$^-$~centers of region A. Coherence times of T$_2^A$=0.66~\textmu s and  T$_2^B$=1.56~\textmu s are measured for the NV$^-$ centers A and B, respectively.}
\label{pic:hahn}
\end{figure}
In order to assess the quality of the produced NV centers, Hahn echo measurements are carried out to determine the coherence time. This coherence time determines the maximum interaction time in potential sensing experiments and therefore plays a crucial role. The results of these measurements for the two $^{15}$NV$^-$ centers of region F can be found in figure \ref{pic:hahn}. It is known that very shallow NV centers exhibit short coherence times \cite{Staudacher2012, Fukuda2018}. Such short coherence stems from the proximity to the diamond surface, where a variety of paramagnetic fluctuators might be present. It is possible to overcome these effects by using differently doped diamond as implantation material \cite{FavarodeOliveira2017} or the recently reported so-called indirect overgrowth technique \cite{Findler2020}. Another explanation for short coherence times might be coupling to highly concentrated neutral-charged substitutional nitrogen (P1 center) in close proximity of the generated NV \cite{Bauch2020}. In our case, due to the tight implantation focus in combination with a low NV creation yield, indeed a high nitrogen atom concentration might be conjectured, as compared with non-focused implantation techniques at high impact energies.

\section{Perspectives}

The demonstrated isotope-selective and maskless deterministic implantation results are encouraging for building a future quantum processor with coupled NV centers in diamond. Even though the yield in this work has been low, in the meantime pre-doping of diamond has been developed that allows for a yield up to 75$\%$~\cite{lue2019}. A pre-implantation with either phosphorous, oxygen or sulfur ions is followed by first annealing, thus preparing for the orders of magnitude improved yield upon the implantation of nitrogen and a second annealing step. Note, that already with a yield exceeding 50$\%$, and in a rectangular lattice of qubits, a central NV center would  have with 94$\%$ chance at least one next-neighbor NV qubit out of  its four closest lattice qubits positions. If a lattice of qubits would be filled in such manner, it would  be sufficiently populated for an effective perculation of entanglement.  Improvements of both, the yield and the coherence properties are expected for shallow NVs if a higher annealing temperature of 1500$\,^{\circ}$C would be used. A second challenge is the selective qubit readout, especially for NV-qubit arrays with a grid dimension of about 10 to 20~nm. This is because optical readout schemes are limited to the Abbe diffraction limitation at about 0.2~$\mu$m, thus the individual NV$^-$ or nuclear spin-qubits can only be distinguished in the frequency domain at cryogenic temperatures, but at the prize of spectral narrowing~\cite{wu2019,abo2019,bradley2019}. A recently demonstrated electrical readout of NV-qubit states~\cite{siy2019} may be a potential solution, as a future architecture for NV-based quantum computing could use nano-wires on top of the diamond surface for selective readout of individual NV qubits at distances of about 20~nm, thus NVs sufficiently close to generate entanglement and execute fast gate operations. 

We thank Kilian Singer, Sam Dawkins, Luc Courturier and Sebastian Wolf for contributions at an earlier stage and acknowledge financial support by the Bundesministerium f\"ur Bildung und Forschung via Q.Link.X., the Volkswagen Stiftung and the Deutsche Forschungsgemeinschaft through the DIP program (grant Schm 1049/7-1).

\section*{References}
\bibliography{literature}

\end{document}